\documentclass[twocolumn,showpacs,preprintnumbers,amsmath,amssymb,prl]{revtex4-1}

\usepackage{graphicx}
\usepackage{dcolumn}
\usepackage{bm}
\usepackage{color}
\usepackage{epsfig}
\usepackage{soul}

\begin{document}

\title{Realization of a double barrier resonant tunneling diode for cavity polaritons}

\author{H. S. Nguyen$^1$}
\author{D.Vishnevsky$^2$}
\author{C. Sturm$^1$}
\author{D. Tanese$^1$}
\author{D. Solnyshkov$^2$}
\author{E. Galopin$^1$}
\author{A. Lema\^itre$^1$}
\author{I. Sagnes$^1$}
\author{A. Amo$^1$}
\author{G. Malpuech$^{2}$}
\author{J. Bloch$^1$}
\email{jacqueline.bloch@lpn.cnrs.fr}
\affiliation{$^1$Laboratoire de Photonique et de Nanostructures, LPN/CNRS, Route de Nozay, 91460 Marcoussis , France}
\affiliation{$^2$Institut Pascal, PHOTON-N2, Clermont Universit\'e, Universit\'e Blaise Pascal, CNRS, 24 avenue des Landais,
63177 Aubi\`ere Cedex, France}
\date{\today}

\begin{abstract}
We report on the realization of a double barrier resonant tunneling diode for cavity polaritons, by lateral patterning of a one-dimensional cavity. Sharp transmission resonances are demonstrated when sending a polariton flow onto the device. We use a non-resonant beam can be used as an optical gate and control the device transmission. Finally we evidence distortion of the transmission profile when going to the high density regime, signature of polariton-polariton interactions.
\end{abstract}

\maketitle

Resonant tunneling diodes (RTD) are primary elements of nanoelectronics providing negative differential resistance and other nonlinear properties \cite{Sollner:1983,RTDbook}. They opened the way for many applications, such as high frequency oscillation \cite{Brown:1991}, resonant tunneling transistor \cite{Capasso:1985} or multiple-valued logic circuits \cite{Mazumder:1998}. Such double barrier structures, when brought to the quantum limit, reveal fascinating  physics, such as the Coulomb blockade which was observed with electrons \cite{Pasquier:1993} and Cooper pairs \cite{Joyez:1994}. Recently, resonant transmission in high quality factor photonic crystal cavities has allowed to evidence bistable behavior and implement optical memories, making use of the carrier-induced non-linearity \cite{Nozaki:2012}. Transport of atomic Bose-Einstein condensates (BEC) through a double barrier structure has also been  theoretically considered by several groups \cite{Carusotto:2000, Paul:2005}, with interesting predictions related to non-linear bosonic interactions. However, no experimental demonstration of these effects for BEC has been reported so far, due to the difficulty of creating adequate potential profile for atomic BEC.   Cavity polaritons have appeared these last years as an alternative system to investigate the physics of out of equilibrium condensates. They are exciton-photon mixed states arising from the strong coupling between the optical mode of a cavity and excitons confined in quantum wells \cite{Weisbuch:1992}. Cavity polaritons propagate with a speed comparable to the speed of light ($\sim$ 1$\%$ speed of light) \cite{Freixanet:2000}  thanks to their photonic component, and simultaneously show strong nonlinear interactions inherited from their excitonic component \cite{Ciuti:2001}. Cavity polaritons are now considered as a new platform for optical devices with many promising proposals for all-optical integrated logical circuits\cite{Liew:2008, Shelykh:2009, Shelykh:2010, Liew:2013}. Experimental demonstrations of a polariton spin switch \cite{AmoNatphot:2010}, polariton transistor \cite{Gao:2012,Ballarini:2012} and of a polariton interferometer \cite{Sturm:2013} have been recently reported . A key advantage of cavity polaritons is that the potential in which they evolve can be engineered at will, either by optical means \cite{WertzNatphys:2010,AmoPRB:2010,Tosi:2012}, by depositing metallic layers on top of the cavity \cite{Lai:2007}, by using a surface acoustic wave \cite{Cerda:2010}, or by etching the cavity into lower dimensionality microstructures \cite{Bloch:1997,Paraiso:2009}.

In this letter, we demonstrate the realization of a polariton RTD based on an innovative design of a wire cavity. Two micron-size constrictions are etched in the wire cavity and create two tunnel barriers, defining an isolated island with discrete confined polariton states. Sending a polariton flow onto this double barrier structure, we observe resonant tunneling when the polariton energy coincides with the energy of one of the confined modes. We show that a non resonant laser beam, focused onto the island, can modulate the RTD transmission with a peak to valley ratio as high as 28. The device is therefore operating as an optical gate. Finally we evidence strong asymmetry in the transmission profiles when going to the high density regime. This is shown to be the signature of non-linear polariton interactions within the island, as fully supported by numerical simulations.

\begin{figure}[htb]
\begin{center}
\includegraphics[width=8.5cm]{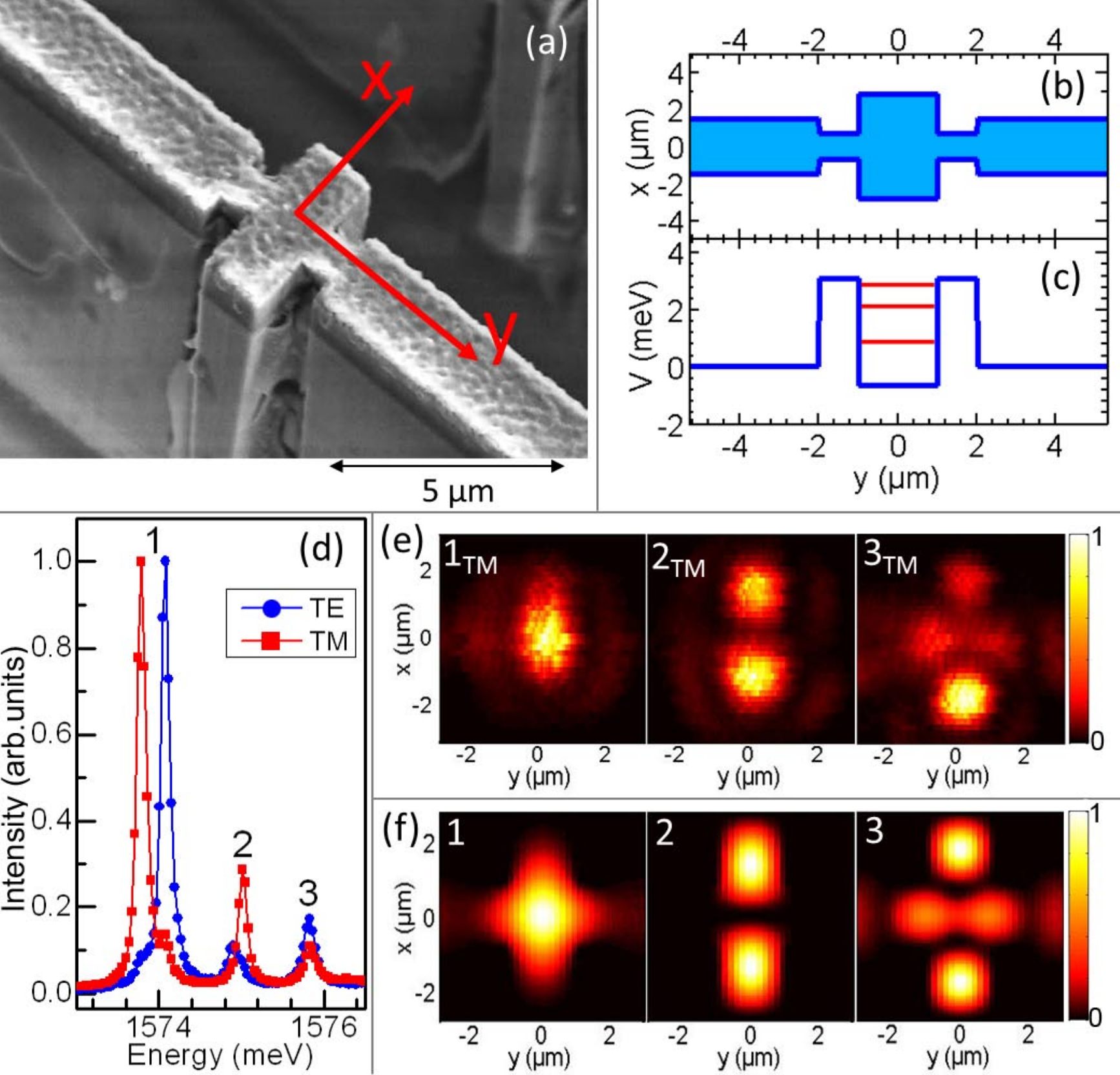}
\end{center}
\caption{\label{fig1}{(a) Scanning electron microscopy image of the polariton RTD. (b) Width profile of the polariton RTD. (c) Simulated potential along the wire in the region of the RTD (blue), and energy of confined states within the island (red). (d) Emission spectra measured on the island for TM (red squares) and TE (blue circles) detection polarization (excitation power $  300$ $\mu$W, laser enery 1.62 eV). (e) Spatially resolved emission of the three lowest energy confined polariton states for TM detection polarization. (f) Calculated emission pattern corresponding to (e).}}
\end{figure}
The sample was grown by molecular beam epitaxy and consists in a $\lambda/2$ microcavity with 28 (resp. 40) pairs of $Ga_{0.8}Al_{0.2}As/Ga_{0.05}Al_{0.95}As$ $\lambda/4$ layers in the top (resp. bottom) distributed Bragg mirror. 12 GaAs quantum wells (7 nm thickness) are inserted in the structures. The quality factor of the microcavity amounts to 100000 and the Rabi splitting  to 15 meV. Electron beam lithography and inductively coupled plasma dry etching were used to fabricate 1D microwires of 3 $\mu$m width and 440 $\mu$m length. The RTD is defined by a microstructure described in Figs.~\ref{fig1}(a) and ~\ref{fig1}(b): two constrictions (width = 1.4 $\mu$m, length = 1 $\mu$m) surround an isolated island (width = 5.6 $\mu$m, length = 2 $\mu$m). As the 1D confinement potential is inversely proportional to the square of the wire width \cite{Dasbach:2002}, the microstructure defines a double potential barrier and a 0D polariton island (see Fig.~\ref{fig1}(c)). Micro-photoluminescence experiments are performed at 10K on single microwires using a cw monomode Ti:sapphire laser, focused onto a 2 $\mu$m spot with a microscope objectif (NA = 0.55). A second cw monomode Ti:sapphire laser is  used for experiments requiring simultaneously both resonant and non-resonant excitation. Polariton emission is imaged on a CCD camera coupled to a monochromator. The studied RTD corresponds to an exciton-photon detuning around -10 meV (defined as the difference between the energy of the photonic and excitonic modes) .
 \begin{figure}[htb]
\begin{center}
\includegraphics[width=8.5cm]{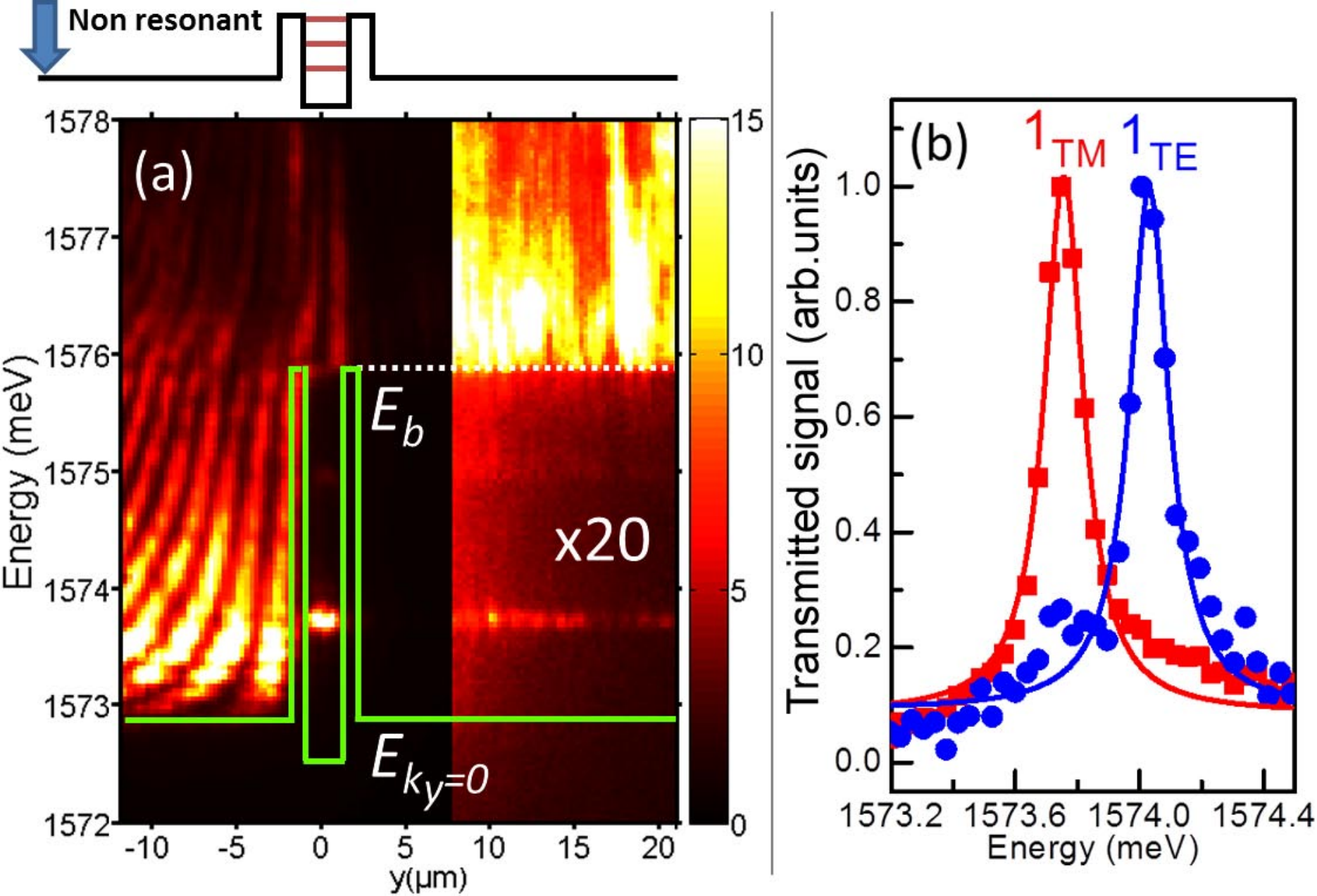}
\end{center}
\caption{\label{fig2}{(a) Spectrally and spatially resolved emission measured under TM detection polarization , when exciting the structure with a non-resonant laser beam of energy $1.62\,eV$ focused at position  y=-50 $\mu$m. (b)Transmission intensity measured under TM (red squares) and TE (blue circles) detection polarization. The signal is integrated between $y=7\,\mu m$ to $y=15\,\mu m$. The red and blue lines are fits with two Lorentzians of 155 $\mu eV$ linewidth, limited by the detection resolution.}}
\end{figure}

We first characterize the polariton modes confined within the island, between the two tunnel barriers. Emission spectra of the island measured under TM (along y-axis) and TE (along x-axis) detection polarization are shown in Fig.~\ref{fig1}(d). These spectra are obtained by exciting the island non-resonantly with excitation power well below condensation threshold. Three discrete polariton modes are observed for each polarization. Polarization splitting reflect the anisotropy of the confinement within the island. Spatial mapping of these confined modes is presented in Fig.~\ref{fig1}(e) for TM polarization. Characteristic emission pattern with well defined emission lobes are observed in good agreement with simulations obtained solving a 2D Schr\"{o}dinger equation for a particle of effective mass  $m = 6.1\,10^{-5}m_{electron}$ in a potential corresponding to that of the heterostructure. These measurements demonstrate that the two constrictions define an isolated island with well defined discrete states.

Let us now discuss polariton transport through this double barrier structure. We first send onto the microstructure a polariton flow with a broad energy distribution. For this experiment, polaritons are injected with non resonant excitation far from the microstructures (typically $50$ to $80\, \mu m$ away). Figure ~\ref{fig2}(a) displays the spectral emission measured along the wire in TM detection polarization. The transmitted signal ($y>3\,\mu m$) presents a sharp threshold energy above which polaritons propagate across the island ($E > E_b = 1575.8$ meV). This energy corresponds to the top of the tunnel barrier, and we deduce a barrier height of $V_b \approx  3.0$ meV. Below $E_b$, the transmission is vanishing except for a few sharp resonances. The lowest one is observed at 1573.7 meV, and corresponds to the resonance of incident polaritons with the  $1_{TM}$ mode of the island. Transmission spectra for energy close to this resonance are reported in  Fig.~\ref{fig2}(b) for both detection polarizations: the observed peaks are splitted by the TE-TM splitting, showing that polariton polarization is preserved in the resonant tunneling. Notice that resonant tunneling transmission corresponding to $2_{TM}$ (i.e. 1575.0 meV) is weak in Fig.~\ref{fig2}(a). This feature can be explained considering the mode mismatch between the 1D incident polariton mode (symmetric along x) and the $2_{TM}$ mode (antisymmetric along x). Finally, notice that the fringe pattern observed in the upstream side ($y<-3\,\mu m$) of Fig.~\ref{fig2}(a) is due to the interferences between incident polaritons and polaritons reflected on the double barrier structure.
\begin{figure}[htb]
\begin{center}
\includegraphics[width=8.5cm]{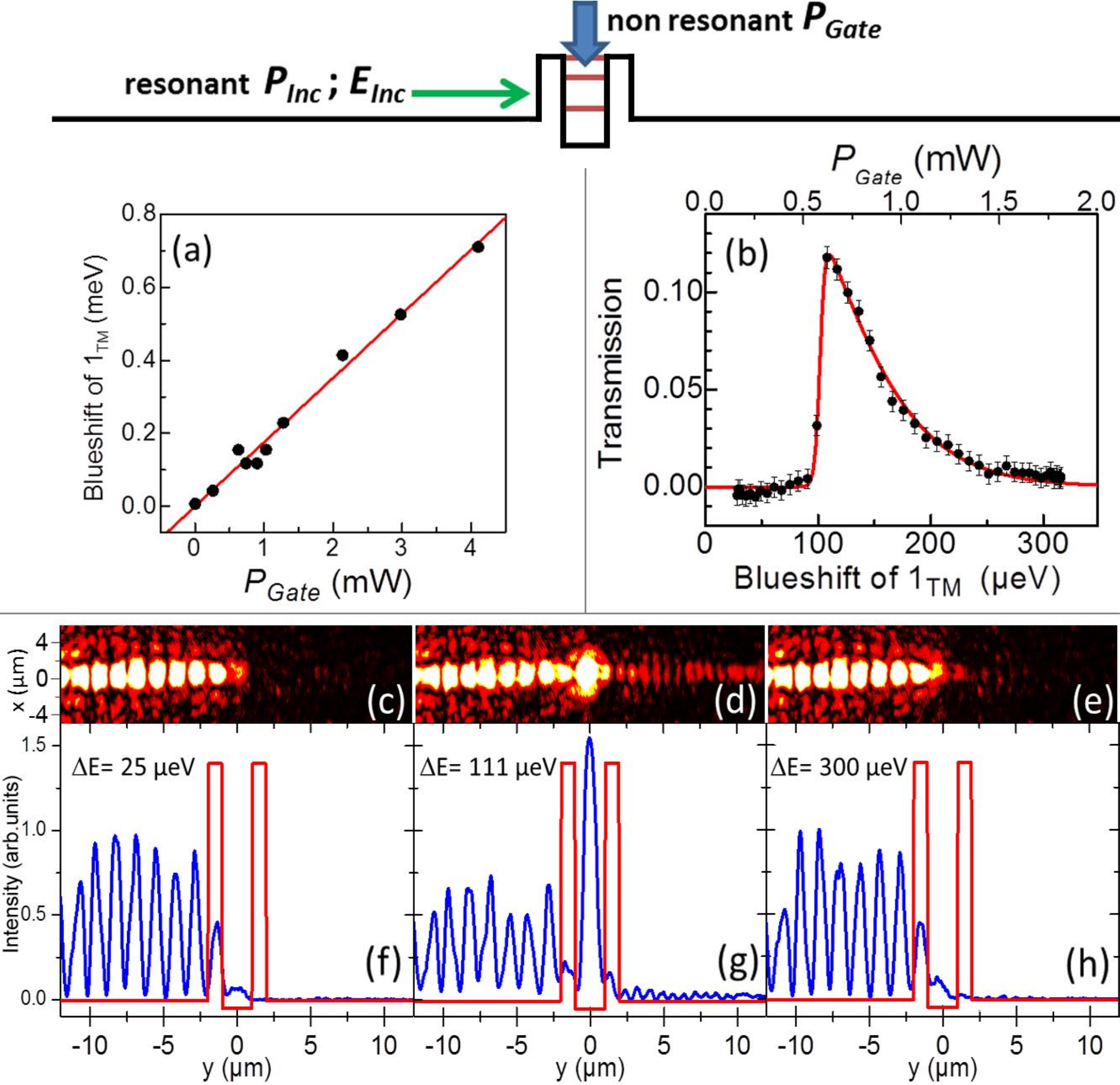}
\end{center}
\caption{\label{fig3}{(a) (black circles) Measured blueshift of the $1_{TM}$ mode as function of the optical gate power $P_{Gate}$. (red line) Linearfit with a $174\,\mu eV. mW^{-1}$ slope. The energy of the laser gate beam is $1.62\,eV$. (b) Tunneling transmission as a function of the blueshift of the $1_{TM}$ mode (bottom axis), or of $P_{Gate}$ (top axis). The red line is a guide to the eye. The polariton flow is injected at $y = - 80\,\mu$m by a resonant laser of power $P_{Inc} = 40\,mW$ and of energy $E_{Inc} = 1573.8\,meV$. (c-e) Spatially resolved emission measured for different values of $P_{Gate}$ corresponding to a blueshift of the $1_{TM}$ mode equal to (c) $25\,\mu eV$, (d) $111\,\mu eV$ and (e) $300\,\mu eV$. (f-h) Corresponding measured integrated intensity as a function of position (blue) and wire potential (red).}}
\end{figure}

In the following, we describe how the transmitted intensity can be optically modulated, using a second laser beam focused on the island. To do this, we use the repulsive interaction of polaritons with a reservoir of excitons \cite{AmoPRB:2010}, which is locally injected in the island using a weak non-resonant optical excitation of power $P_{Gate}$. When $P_{Gate}$ is turned on, polariton-exciton interactions induce a $174\,\mu eV.mW^{-1}$ blueshift of the polariton modes confined within the island (see Figure~\ref{fig3}(a)). We use this gate beam to control the transport of a monochromatic TM polarized polariton flow, which is sent onto the double barrier structure, using a resonant laser beam (of energy $E_{Inc}$ and power $P_{Inc}$). We chose $E_{Inc} = 1573.8\,meV$, a value  slightly larger than the energy of the $1_{TM}$  mode (which lies at 1573.7 meV). Figure~\ref{fig3}(b) reports the measured transmission for different values of $P_{Gate}$.  When the gate beam brings the energy of the $1_{TM}$ mode into resonance with the polariton flow, a pronounced increase in the transmission is induced. This is further illustrated in Fig.~\ref{fig3}(c-h), where spatially resolved emission is monitored for values of $P_{Gate}$ corresponding to energy of the $1_{TM}$ mode below, at and above the resonance. We clearly observed an enhanced luminescence signal in the downstream region at resonance (see Fig.~\ref{fig3}(d,g)), which corresponds to an induced blueshift of 111 $\mu$eV. On the opposite, vanishing transmitted signal is observed for a blueshift of $25\,\mu$eV or  $300\,\mu$eV. These results prove that our device is indeed operated as an all-optically-controlled RTD, with a very high spectral selectivity. The peak (valley) transmission amounts to $11.1\,\%$ ($0.4\%$) resulting in a Peak-To-Valley Signal Ratio (PVSR) of $28$ \cite{Supp}.

An interesting feature is revealed in the transmission spectrum presented in Fig.~\ref{fig3}(b): the line-shape is distorted with respect to a Lorentzian profile, with a more abrupt shape on the low power side. We show below that this is a direct evidence of polariton-polariton interactions \cite{Ciuti:2001} within the island, resulting in a nonlinear tunneling regime of the device. Let us first give a qualitative explanation of the observed asymmetric transmission profile. When $P_{Gate}$ is slightly lower than the resonance power $P_{res}$, incident polaritons start to enter the island. As a consequence, the energy of $1_{TM}$ undergoes an additional blueshift due to the interactions between polaritons confined in the island, which reduces the detuning between $E_{Inc}$ and the energy of $1_{TM}$. This mechanism thus provides a positive feedback that accelerates the passage to the resonant tunneling. On the contrary, when $P_{Gate}$ slightly exceeds $P_{res}$, less polaritons enter the island. Thus, the additional blueshift decreases and the passage to off-resonant tunneling is slowed down due to a negative feedback.

\begin{figure}[htb]
\begin{center}
\includegraphics[width=8cm]{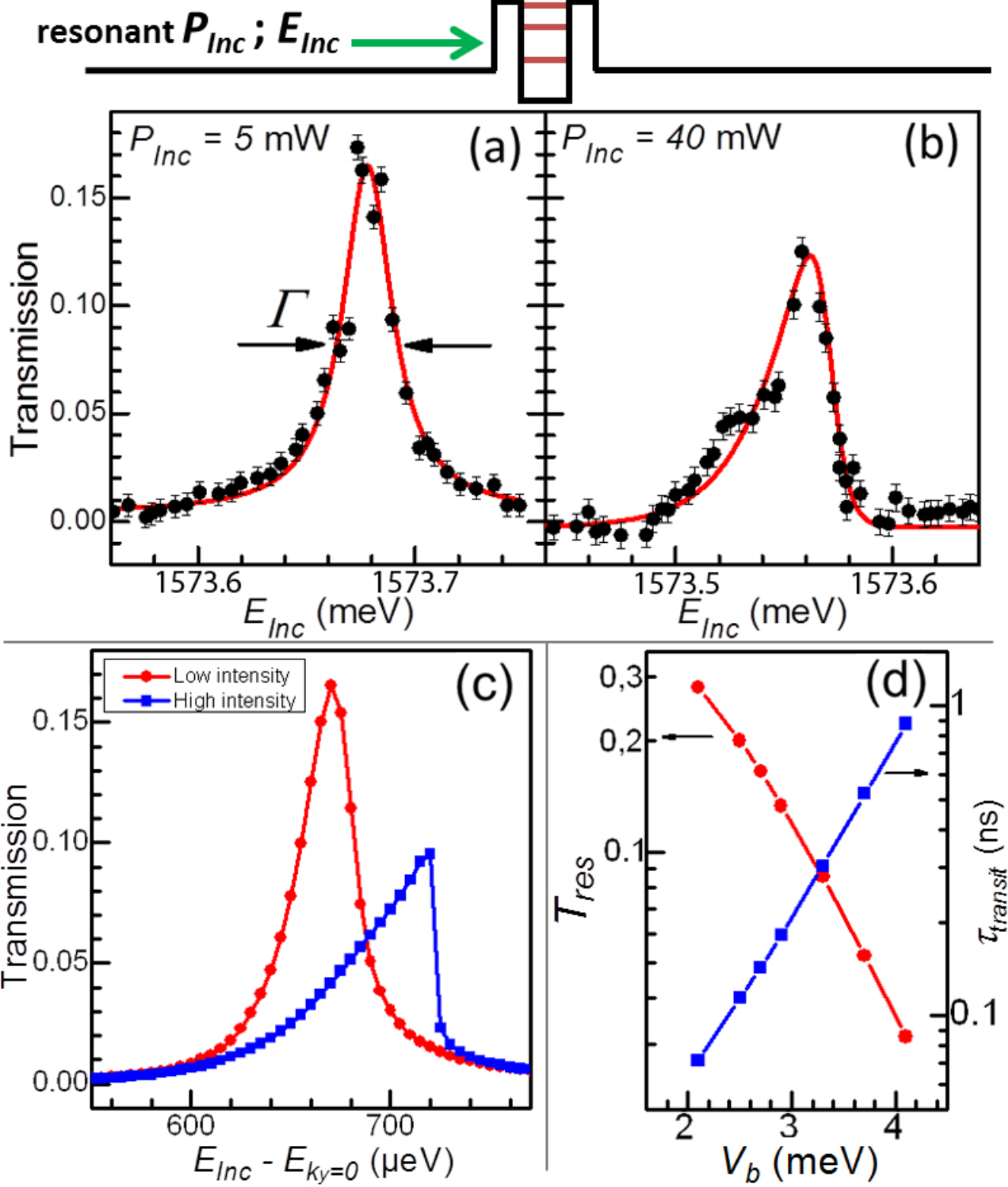}
\end{center}
\caption{\label{fig4}{(a,b) Tunneling transmission measured as a function of $E_{Inc}$ for (a) $P_{Inc} = 5 \,mW$ and (b) $P_{Inc} = 40 \,mW$. The red line in (a) is a Lorentzian fit of linewidth $27\, \mu eV$, while the red line in (b) is a guide to the eye. The polariton flow is injected at $y= -80$ $\mu m$. (c) Simulation of the resonant tunneling transmission profile for low pump density (red circles) and high pump density (blue squares). Simulation parameters: polariton radiative lifetime $\tau_{rad}= 34$ ps, barrier potential $V_b = 2.7 \,meV$, interaction constant $g=0.4\,\mu eV. \mu m$, incident polariton density  $65 \,\mu m^{-1}$  (resp. $390\,\mu m^{-1}$) for the low (resp. high) pump density case. (d) Simulation of the resonant transmission $T_{res}$ (red circles) and the tunneling transit time $\tau_{transit}$ (blue squares) as a function of the barrier height.}}
\end{figure}

Of course this non-linear regime occurs when the polariton density of the incident flow is large enough. If we reduce $P_{Inc}$ sufficiently to enter the linear regime, then the luminescence from the island induced by $P_{Gate}$ starts to dominate the emission spectra, and precise extraction of the transmission profiles becomes delicate. To unambiguously demonstrate the two regimes (linear and nonlinear), we performed experiments with a single laser beam (the one injecting the polariton flow)  and probe the transmission profile when scanning the incident energy $E_{Inc}$. For low incident power, a symmetrical transmission spectrum is measured characteristic of the linear regime (see Figure~\ref{fig4}(a) corresponding to $P_{Inc}\,=5\,mW$). The  transmission profile is well fitted by a Lorentzian of linewidth $\Gamma = 27$ $\mu\,eV$, attributed to the $1_{TM}$ mode homogeneous linewidth. In analogy to the non dissipative case, where the tunneling transit time $\tau_{transit}$\cite{Buttinker:1982} is twice the particle lifetime in the island\cite{Peskin:2002}, we can write here: $\Gamma = 2\hbar/\tau_{transit} + \hbar/\tau_{rad}$, where $\tau_{rad}$ is the polariton radiative lifetime within the island, governed by the escape of the photon through the mirrors or the sidewalls the structure. Moreover the value of the transmission peak is given by  $T_{res}\,=\tau_{rad}/(\tau_{rad}+\tau_{transit})\approx 0.165$. Thus we deduce $\tau_{transit} = 172\,ps$  and $\tau_{rad}= 34\,ps$. Notice that $\tau_{rad}$ is close to the nominal photon lifetime calculated from the quality factor of the non-etched structure, expected to be around $40\,ps$.  For larger  $P_{Inc}$, asymmetry of the transmission spectrum develops as illustrated in Fig~\ref{fig4}(b) for $P_{Inc}\,=40\,mW$. This asymmetrical profile, induced by polariton-polariton interactions, is a mirror image of the one shown in Fig.~\ref{fig3}(b), as here $E_{Inc}$ is scanned with respect to the energy of the  $1_{TM}$ mode, whereas before the energy of $1_{TM}$ was tuned with respect to  $E_{Inc}$ using the gate beam. Such asymmetric shape of transmission profile due to nonlinear interaction has already been predicted for cold atom condensate transport through a double potential barrier \cite{Paul:2005}, and also been observed for resonant tunneling of guided light propagation through a photonic crystal microcavity \cite{Nozaki:2012}. To describe our experiments, we developed numerical simulations  taking into account not only polariton-polariton interactions, but also the finite polariton lifetime, which is the specificity of our dissipative system.  Figure.~\ref{fig4}(c) presents the numerical simulations of transmission profile in the linear and nonlinear regime obtained with an interaction constant $g=0.4\,\mu eV. \mu m$ and a polariton density of $65  $ $\mu m^{-1}$ and $390 $ $\mu m^{-1}$. A 1D Gross-Pitaevskii equation has been used for  Fig.~\ref{fig4}(c), and the validity of this approximation has been confirmed by 2D simulations. The calculated profiles reproduce the observed features. However, while the simulation shown in Fig.~\ref{fig4}(c) and the theoretical calculation for atom condensates \cite{Paul:2005} predict a blueshift of the resonant tunneling peak in the nonlinear regime with respect to the one in the linear regime, a redshift is observed in our experimental results (see Figs.~\ref{fig4}(a) and .~\ref{fig4}(b)). We think that this discrepancy between theory and experiment is due local heating of the sample in the high density regime.

Finally we would like to discuss the performance of our polariton RTD device.  In the present experiment, the peak transmission is limited to $16.5 \%$ because of the long tunneling time as compared to the photon lifetime.  Further engineering of the tunnel barrier could allow controlling the tunneling transit time and thus optimizing the transmission. Indeed, Fig.~\ref{fig4}(d) presents simulation of the resonant transmission and the tunneling transit time corresponding to different barrier heights. Our simulation points that resonant transmission coefficient can be significantly increased when reducing the barrier height. Concerning the dynamics of our RTD device, the optical gating dynamic is expected to be limited by the lifetime of the excitonic reservoir, which is around 400 ps \cite{Bajoni:2006}. Thus, we estimate that our device could operate at frequency of several GHz. We could envisage to increase this operating speed by using a resonant optical gate. Indeed then the dynamics would be mainly governed by the polariton lifetime which is much shorter.

In conclusion, we have fabricated a polariton RTD exhibiting resonant polariton transport through a double potential barrier structure. The device is gated by a low-power non-resonant laser, which modulates the transmission within a PVSR of 28. Nonlinear transmission regime of a BEC is demonstrated experimentally for the first time.  These results open the way for a new generation of  integrated logical circuit by exploiting for instance the nonlinear transport of many RTDs in a more complex architecture \cite{Liew:2013}. Moreover, by reducing the size of the isolated island, the quantum regime could be reached: resonant tunneling transmission could become sensitive to single-polariton non-linearity \cite{Verger:2006}. In this regime, emission of non classical light is expected because of  polariton blockade, together with many fascinating features of the Bose-Hubbard physics \cite{Liew:2010, Bamba:2011,Greentree:2006,Hartmann:2006}.

We thank Luc Le Gratiet for scanning electron microscopy. The work was supported financially by the 'Agence Nationale pour la Recherche' project "Perocai" and "Quandyde" (ANR-11-BS10-001), by the FP7 ITN "Clermont4" (235114) and "Spin-Optronics" (237252), by the french RENATECH network and the RTRA "Triangle de la Physique" contract "Boseflow1D" and "2012-0397-Interpol".

\end{document}